\documentclass[ amsmath, amssymb, aps, pra, 10pt, twocolumn, groupedaddress, nofootinbib,
    longbibliography, superscriptaddress ]{revtex4-1}

\usepackage[T1]{fontenc} 
\usepackage[utf8]{inputenc}

\usepackage{graphicx, epsf} 
\usepackage{color} 
\usepackage[table]{xcolor}
\usepackage{dcolumn} 
\usepackage{import} 
\usepackage[skins]{tcolorbox} 
\usepackage{titlesec, setspace} 
\usepackage{listings} 
\usepackage{caption} 
\usepackage[bitstream-charter]{mathdesign} 
\graphicspath{{figures/}}
\input{Qcircuit}
\usepackage{algorithm}
\usepackage[noend]{algpseudocode}

\usepackage{calrsfs}
\DeclareMathAlphabet{\pazocal}{OMS}{zplm}{m}{n} \let\mathcal\undefined
\newcommand{\mathcal}[1]{\pazocal{#1}}

\usepackage[newfloat,frozencache,cachedir=minted]{minted}

\usepackage{tabularx,booktabs}
\usepackage{cellspace}
\usepackage{placeins}

\newcolumntype{L}[1]{>{\raggedright\arraybackslash}p{#1}}
\arrayrulecolor{white}
\usepackage[hyperfootnotes=true,colorlinks=true,allcolors=blue]{hyperref}
\newtheorem{defn}{Definition}

\newcommand{\node}[2][]{{\begin{array}{c} \ _{#1}\  \\ {#2} \\ \ \end{array}}\drop\frm{o} }

\newcommand{\inp}{x}
\newcommand{\vars}{\theta}

\newcommand{\expval}[1]{\langle#1\rangle}
\newcommand{\Var}[1]{\text{Var}(#1)} \newcommand{\estimator}[1]{#1^*}

\definecolor{lightgreen}{HTML}{19b37b}
\definecolor{xgreen}{HTML}{19b37b}

\definecolor{codegreen}{rgb}{0,0.6,0}
\definecolor{codegray}{rgb}{0.5,0.5,0.5}
\definecolor{codepurple}{rgb}{0.58,0,0.82}
\definecolor{backcolour}{RGB}{230, 240, 230}
\definecolor{BlueChill}{RGB}{32, 147, 149}
\definecolor{Shamrock}{RGB}{50, 211, 167}

\titleformat{\section}[display]{\vspace{-1em}}{}{0pt}{\normalfont\sffamily\color{lightgreen}}[\vspace{-3pt}\hrule]
\titleformat{\subsection}[display]{\vspace{-2.5em}}{}{0pt}{\normalfont\sffamily\color{lightgreen}}[\vspace{-1em}]
\titleformat{\subsubsection}[display]{\vspace{-2.5em}}{}{0pt}{\bfseries\color{black}}[\vspace{-1em}]

\makeatletter

\def\frontmatter@abstractfont{\sffamily\color{black}\setstretch{1.5}}%
\def\frontmatter@title@format{\noindent\huge\sffamily\color{lightgreen}}{}%
\def\frontmatter@authorformat{\vspace{1em}\noindent\color{lightgreen}\Large\sffamily}%
\def\frontmatter@affiliationfont{\vspace{1em}\color{lightgreen}\noindent\normalsize\sffamily}%
\def\frontmatter@above@affiliation@script{\vspace{1em}\noindent}%
\def\frontmatter@makefnmark{} \renewcommand*\frontmatter@date[2][\Dated@name]{\def\@date{}}%
\makeatother

\titleformat{\section}[display]{\vspace{-1em}}{}{0pt}{\Large}[\vspace{-3pt}\color{xgreen}]
\titleformat{\subsection}[display]{\vspace{-2.5em}}{}{0pt}{\bfseries\color{codegray}}[\vspace{-1em}]
\titleformat{\subsubsection}[display]{\vspace{-2.5em}}{}{0pt}{\itshape\color{black}}[\vspace{-1em}]

\makeatletter
\def\frontmatter@abstractfont{\color{black}\setstretch{1.2}}%
\def\frontmatter@title@format{\noindent\huge\color{black}}{}%
\def\frontmatter@authorformat{\vspace{1em}\noindent\color{codegray}\normalsize}%
\def\frontmatter@affiliationfont{\itshape\color{black}\noindent\footnotesize}%
\def\frontmatter@above@affiliation@script{\vspace{1em}\noindent}%
\def\frontmatter@makefnmark{} \renewcommand*\frontmatter@date[2][\Dated@name]{\def\@date{}}%
\makeatother

\usepackage[format=plain,labelfont={bf,it},textfont=it]{caption}
\newenvironment{code}{\captionsetup{type=listing}}{}
\SetupFloatingEnvironment{listing}{name=Codeblock}
\setminted[python]{ bgcolor=gray!6, fontfamily=tt, gobble=0, fontsize=\footnotesize,
breaklines=true,
    framerule=0.4pt,
    funcnamehighlighting=true, tabsize=2, obeytabs=false, mathescape=false samepage=true,
    showspaces=false, showtabs =false, texcl=false, linenos=false, xleftmargin=\parindent,
    numbersep=1pt, texcomments }

\newmintedfile[pythonfile]{python}{}

\usetikzlibrary{calc}
\usetikzlibrary{shapes.multipart}
\definecolor{quantum1}{HTML}{8EDBCE}
\definecolor{quantum2}{HTML}{3F605B}
\definecolor{exp1}{HTML}{9AB9ED}
\definecolor{exp2}{HTML}{204177}
\definecolor{classical}{HTML}{EBBA92}
\tikzset{input node/.style={}} \tikzset{quantum node/.style={draw, align=center, anchor=west, inner
sep=5pt,rounded corners=4pt, rectangle split, rectangle split horizontal, rectangle split parts=2,
rectangle split part fill={quantum1,quantum2}, every two node part/.style={text = white}}}
\tikzset{classical node/.style={draw, rectangle,align=center, anchor=west, thin, fill=classical,
inner sep=5pt}} \tikzset{output node/.style={}} \tikzset{out label/.style={midway, above}}
\tikzset{connector/.style={anchor=center, opacity=0.}} \tikzset{samples label/.style={at start,
below, xshift=4pt}}

\usepackage{float}
\floatstyle{ruled}
\newfloat{floatbox}{htp}{box}
\floatname{floatbox}{Box}
\newcommand{\infobox}[2]{
    \begin{floatbox}
        \caption{#1}
        \let\centering\relax
        \parbox{.94\columnwidth}{#2}
    \end{floatbox}
}

\usepackage{silence}
\begin{document}
\title{PennyLane: Automatic differentiation of hybrid quantum-classical computations}

\begin{abstract}
PennyLane is a Python~3 software framework for differentiable programming of quantum computers. The
library provides a unified architecture for near-term quantum computing devices, supporting both
qubit and continuous-variable paradigms. PennyLane's core feature is the ability to compute
gradients of variational quantum circuits in a way that is compatible with classical techniques such
as backpropagation. PennyLane thus extends the automatic differentiation algorithms common in
optimization and machine learning to include quantum and hybrid computations. A plugin system makes
the framework compatible with any gate-based quantum simulator or hardware. We provide plugins for
hardware providers including the Xanadu Cloud, Amazon Braket, and IBM Quantum, allowing
PennyLane optimizations to be run on publicly accessible quantum devices. On the classical front,
PennyLane interfaces with accelerated machine learning libraries such as TensorFlow, PyTorch, JAX,
and Autograd. PennyLane can be used for the optimization of variational quantum eigensolvers,
quantum approximate optimization, quantum machine learning models, and many other applications.
\end{abstract}

\author{Ville Bergholm}
\affiliation{\textsuperscript{1}Xanadu, 777 Bay Street, Toronto, Canada}
\author{Josh Izaac}
\affiliation{\textsuperscript{1}Xanadu, 777 Bay Street, Toronto, Canada}
\author{Maria Schuld}
\affiliation{\textsuperscript{1}Xanadu, 777 Bay Street, Toronto, Canada}
\author{Christian Gogolin}
\affiliation{\textsuperscript{1}Xanadu, 777 Bay Street, Toronto, Canada}

\author{Shahnawaz Ahmed}
\affiliation{\textsuperscript{2}Wallenberg Centre for Quantum Technology, Department of
Microtechnology and Nanoscience, Chalmers University of Technology, 412 96 Gothenburg, Sweden}

\author{Vishnu Ajith}
\affiliation{\textsuperscript{3}Indian Institute of Information Technology, Kottayam}

\author{M. Sohaib Alam}
\affiliation{\textsuperscript{4}Quantum Artificial Intelligence Laboratory (QuAIL), NASA Ames Research Center, Moffett Field, CA, 94035, USA}
\affiliation{\textsuperscript{5}USRA Research Institute for Advanced Computer Science (RIACS), Mountain View, CA, 94043, USA}

\author{Guillermo Alonso-Linaje}
\affiliation{\textsuperscript{1}Xanadu, 777 Bay Street, Toronto, Canada}

\author{B. AkashNarayanan}
\noaffiliation

\author{Ali Asadi}
\affiliation{\textsuperscript{1}Xanadu, 777 Bay Street, Toronto, Canada}

\author{Juan Miguel Arrazola}
\affiliation{\textsuperscript{1}Xanadu, 777 Bay Street, Toronto, Canada}

\author{Utkarsh Azad}
\affiliation{\textsuperscript{1}Xanadu, 777 Bay Street, Toronto, Canada}

\author{Sam Banning}
\affiliation{\textsuperscript{1}Xanadu, 777 Bay Street, Toronto, Canada}

\author{Carsten Blank}
\affiliation{\textsuperscript{6}data cybernetics,  Martin-Kolmsperger-Str 26, 86899 Landsberg,
Germany}

\author{Thomas R Bromley}
\affiliation{\textsuperscript{1}Xanadu, 777 Bay Street, Toronto, Canada}

\author{Benjamin A. Cordier}
\affiliation{\textsuperscript{7}Department of Medical Informatics and Clinical Epidemiology, Oregon Health and Science University, Portland, OR 97202, USA}

\author{Jack Ceroni}
\affiliation{\textsuperscript{1}Xanadu, 777 Bay Street, Toronto, Canada}

\author{Alain Delgado}
\affiliation{\textsuperscript{1}Xanadu, 777 Bay Street, Toronto, Canada}

\author{Olivia Di Matteo}
\affiliation{\textsuperscript{1}Xanadu, 777 Bay Street, Toronto, Canada}
\affiliation{\textsuperscript{8}Dept. of Electrical and Computer Engineering, The University of British Columbia, Vancouver, BC, V6T 1Z4, Canada}

\author{Amintor Dusko}
\affiliation{\textsuperscript{1}Xanadu, 777 Bay Street, Toronto, Canada}

\author{Tanya Garg}
\affiliation{\textsuperscript{9}Department of Physics, Indian Institute of Technology Roorkee, Roorkee, Uttarakhand, India}

\author{Diego Guala}
\affiliation{\textsuperscript{1}Xanadu, 777 Bay Street, Toronto, Canada}

\author{Anthony Hayes}
\affiliation{\textsuperscript{1}Xanadu, 777 Bay Street, Toronto, Canada}

\author{Ryan Hill}
\affiliation{\textsuperscript{10}qBraid, 5235 South Harper Court, Chicago, IL 60615}

\author{Aroosa Ijaz}
\affiliation{\textsuperscript{1}Xanadu, 777 Bay Street, Toronto, Canada}

\author{Theodor Isacsson}
\affiliation{\textsuperscript{1}Xanadu, 777 Bay Street, Toronto, Canada}

\author{David Ittah}
\affiliation{\textsuperscript{1}Xanadu, 777 Bay Street, Toronto, Canada}

\author{Soran Jahangiri}
\affiliation{\textsuperscript{1}Xanadu, 777 Bay Street, Toronto, Canada}

\author{Prateek Jain}
\affiliation{\textsuperscript{11}Factal Analytics, Level 2 Chimes Building Plot 61,
Sector - 44, Gurgaon 122003, Haryana, India}

\author{Edward Jiang}
\affiliation{\textsuperscript{1}Xanadu, 777 Bay Street, Toronto, Canada}

\author{Ankit Khandelwal}
\affiliation{\textsuperscript{12}Centre for High Energy Physics, Indian Institute of Science, Bengaluru, Karnataka, India 560012}

\author{Korbinian Kottmann}
\affiliation{\textsuperscript{13}ICFO - Institut de Ciencies Fotoniques, The Barcelona Institute of Science and Technology,
Av. Carl Friedrich Gauss 3, 08860 Castelldefels (Barcelona), Spain}

\author{Robert A. Lang}
\affiliation{\textsuperscript{14}Chemical Physics Theory Group, Department of Chemistry, University of Toronto, Canada}

\author{Christina Lee}
\affiliation{\textsuperscript{1}Xanadu, 777 Bay Street, Toronto, Canada}

\author{Thomas Loke}
\affiliation{\textsuperscript{15}DUG Technology, 76 Kings Park Rd, West Perth WA 6005 Australia}

\author{Angus Lowe}
\affiliation{\textsuperscript{1}Xanadu, 777 Bay Street, Toronto, Canada}

\author{Keri McKiernan}
\affiliation{\textsuperscript{16}Rigetti Computing, 2919 Seventh Street, Berkeley, CA 94710}

\author{Johannes Jakob Meyer}
\affiliation{\textsuperscript{17}Dahlem Center for Complex Quantum Systems, Freie Universit\"at
Berlin, 14195 Berlin, Germany}

\author{J. A. Monta\~nez-Barrera}
\affiliation{\textsuperscript{18}Institute for Advanced Simulation, Jülich Supercomputing Centre, Forschungszentrum Jülich, 52425 Jülich, Germany}

\author{Romain Moyard}
\affiliation{\textsuperscript{1}Xanadu, 777 Bay Street, Toronto, Canada}

\author{Zeyue Niu}
\affiliation{\textsuperscript{1}Xanadu, 777 Bay Street, Toronto, Canada}

\author{Lee James O'Riordan}
\affiliation{\textsuperscript{1}Xanadu, 777 Bay Street, Toronto, Canada}

\author{Steven Oud}
\affiliation{\textsuperscript{19}University of Amsterdam}

\author{Ashish Panigrahi}
\affiliation{\textsuperscript{20}School of Physical Sciences, National Institute of Science Education and Research, HBNI, Jatni, 752 050 Odisha, India}

\author{Chae-Yeun Park}
\affiliation{\textsuperscript{1}Xanadu, 777 Bay Street, Toronto, Canada}

\author{Daniel Polatajko}
\affiliation{\textsuperscript{21}Cervest}

\author{Nicol\'as Quesada}
\affiliation{\textsuperscript{1}Xanadu, 777 Bay Street, Toronto, Canada}

\author{Chase Roberts}
\affiliation{\textsuperscript{1}Xanadu, 777 Bay Street, Toronto, Canada}

\author{Nahum S\'a}
\affiliation{\textsuperscript{22}Centro Brasileiro de Pesquisas Físicas, Rua Dr. Xavier Sigaud 150, 22290-180 Rio de Janeiro, Brazil}

\author{Isidor Schoch}
\affiliation{\textsuperscript{23}ETH Zurich, Quantum Engineering, Department of Information Technology and Electrical Engineering (D-ITET), 8092 Zurich, Switzerland.}

\author{Borun Shi}
\affiliation{\textsuperscript{24}Neo4j UK Ltd.}

\author{Shuli Shu}
\affiliation{\textsuperscript{1}Xanadu, 777 Bay Street, Toronto, Canada}

\author{Sukin Sim}
\affiliation{\textsuperscript{25}Zapata Computing, Inc.}

\author{Arshpreet Singh}
\affiliation{\textsuperscript{26}ITC Infotech Bangalore}

\author{Ingrid Strandberg}
\affiliation{\textsuperscript{27}Department of Microtechnology and Nanoscience MC2, Chalmers University of Technology, SE-412 96 G \"oteborg, Sweden}

\author{Jay Soni}
\affiliation{\textsuperscript{1}Xanadu, 777 Bay Street, Toronto, Canada}

\author{Antal Sz\'ava}
\affiliation{\textsuperscript{1}Xanadu, 777 Bay Street, Toronto, Canada}

\author{Slimane Thabet}
\affiliation{\textsuperscript{28}Pasqal, 7 rue Léonard de Vinci, 91300 Massy, France}
\affiliation{\textsuperscript{29}LIP6, CNRS, Sorbonne Universit\'e, 4 place Jussieu, 75005 Paris, France}

\author{Rodrigo A. Vargas-Hern\'andez}
\affiliation{\textsuperscript{14}Chemical Physics Theory Group, Department of Chemistry, University of Toronto, Canada}
\affiliation{\textsuperscript{30}Vector Institute, MaRS Centre, Toronto, Ontario, M5G 1M1, Canada}

\author{Trevor Vincent}
\affiliation{\textsuperscript{1}Xanadu, 777 Bay Street, Toronto, Canada}

\author{Nicola Vitucci}
\noaffiliation

\author{Maurice Weber}
\affiliation{\textsuperscript{31}ETH Z\"urich, Department of Computer Science, 8092 Z\"irich, Switzerland.}

\author{David Wierichs}
\affiliation{\textsuperscript{32}Institute for Theoretical Physics, University of Cologne, Germany}

\author{Roeland Wiersema}
\affiliation{\textsuperscript{30}Vector Institute, MaRS Centre, Toronto, Ontario, M5G 1M1, Canada}
\affiliation{\textsuperscript{33}Department of Physics and Astronomy, University of Waterloo, Ontario, N2L 3G1, Canada}

\author{Moritz Willmann}
\noaffiliation

\author{Vincent Wong}
\affiliation{\textsuperscript{34}TRIUMF, Vancouver, BC, Canada V6T 2A3}

\author{Shaoming Zhang}
\affiliation{\textsuperscript{35}Technical University of Munich, Department of Informatics, Boltzmannstraße 3, 85748 Garching, Germany}
\affiliation{\textsuperscript{36}BMW Group, Munich, Germany}

\author{Nathan Killoran}
\affiliation{\textsuperscript{1}Xanadu, 777 Bay Street, Toronto, Canada}
\maketitle

\section{Introduction}

Recent progress in the development and commercialization of quantum technologies has had a profound
impact on the landscape of quantum algorithms. Near-term quantum devices require routines that are
of shallow depth and robust against errors. The design paradigm of \emph{hybrid algorithms} which
integrate quantum and classical processing has therefore become increasingly important. Possibly the
most well-known class of hybrid algorithms is that of \emph{variational circuits}, which are
parameter-dependent quantum circuits that can be optimized by a classical computer with regards to a
given objective. \\

Hybrid optimization with variational circuits opens up a number of new research avenues for
near-term quantum computing with applications in quantum chemistry \cite{peruzzo2014variational},
quantum optimization \cite{farhi2014quantum}, factoring \cite{anschuetz2018variational}, state
diagonalization \cite{larose2018variational}, and quantum machine learning~\cite{romero2017quantum,
johnson2017qvector, verdon2017quantum, farhi2018classification, schuld2018quantum,
mitarai2018quantum, schuld2018circuit, grant2018hierarchical, liu2018differentiable,
dallaire2018quantum, havlicek2018supervised, chen2018universal, killoran2018continuous,
steinbrecher2018quantum}. In a reversal from the usual practices in quantum computing research, a
lot of research for these mostly heuristic algorithms necessarily focuses on numerical experiments
rather than rigorous mathematical analysis. Luckily, there are various publicly accessible platforms
to simulate quantum algorithms \cite{wecker14liquid, smelyanskiy2016qhipster, ibmqiskit,
steiger2018projectq, rigettiforest, microsoftqdk, killoran2018strawberry, googlecirq} or even run
them on real quantum devices through a cloud service \cite{arrazola2021quantum, braket, ibmq}. Prior
to PennyLane's launch in 2018, while some frameworks were designed with variational circuits in mind
\cite{killoran2018strawberry, qmlt, sim2018framework}, this was not the norm, and there was at this
stage no unified tool for the hybrid optimization of quantum circuits across quantum platforms,
allowing integration with machine learning tooling while treating all quantum devices on the same
footing\footnote{Since PennyLane was released, other differentiable hybrid optimization frameworks
have followed suit, including TensorFlow Quantum \cite{broughton2020tensorflow} and Yao.jl
\cite{luo2020yao}.}.

PennyLane is an open-source Python~3 framework that facilitates the optimization of quantum and
hybrid quantum-classical algorithms through differentiable quantum programming. It extends several
seminal machine learning libraries --- including \emph{Autograd}~\cite{maclaurin2015autograd},
\emph{TensorFlow}~\cite{abadi2016tensorflow}, \emph{PyTorch} \cite{paszke2017automatic}, and
\emph{JAX} \cite{jax2018github} --- to handle modules of quantum information processing. This can be
used to optimize  variational quantum circuits in applications such as \emph{quantum approximate
optimization}~\cite{farhi2014quantum} or \emph{variational quantum
eigensolvers}~\cite{peruzzo2014variational}. The framework can also handle more complex machine
learning tasks such as training a hybrid quantum-classical machine learning model in a supervised
fashion, or training a generative adverserial network, both when discriminator and generator are
quantum models \cite{dallaire2018quantum} and when one is quantum and the other is classical
\cite{lloyd2018quantum}. Finally, PennyLane introduces the concept of differentiable quantum
transforms --- the ability to map between circuits and their intermediate classical processing steps
in a differentiable manner, as used in many quantum subroutines \cite{di2022quantum}. This enables a
fully differentiable quantum programming paradigm, where the model (the sequence of quantum
transforms) can be optimized alongside the quantum circuits.

\begin{figure}[bh]
\begin{center}
\includegraphics[width=\linewidth]{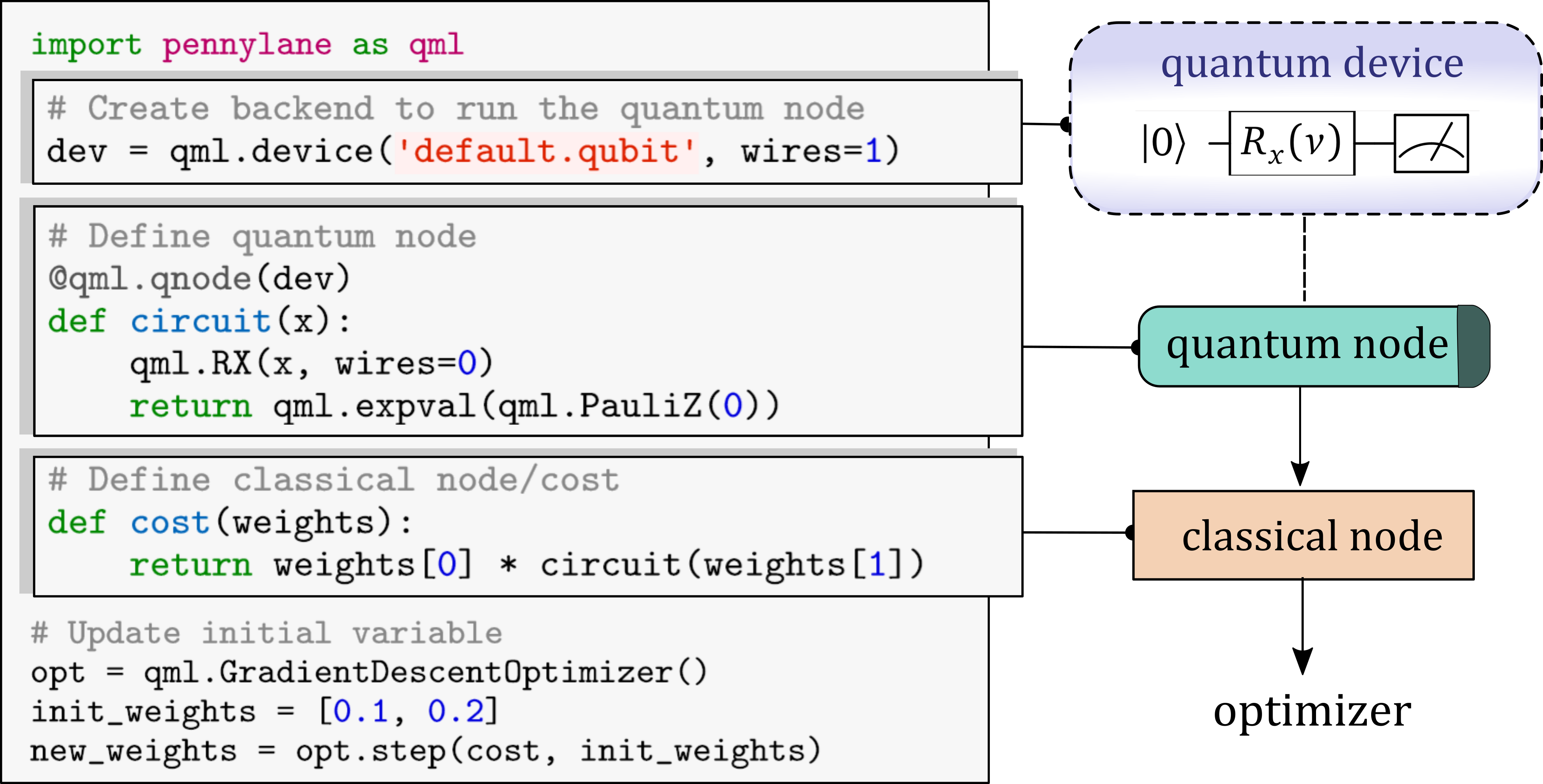}
\end{center}
\caption{
    Basic example of a PennyLane program consisting of a quantum node followed
    by a classical node. The output of the classical node is the objective for optimization.
}
\label{Fig:toy}
\end{figure}

PennyLane can in principle be used with any gate-based quantum computing platform as a backend,
including both qubit and continuous-variable architectures, and has a simple Python-based user
interface. Fig.~\ref{Fig:toy} shows an example that illustrates the core idea of the
framework. The user defines a quantum circuit in the quantum function \verb+circuit+ connected to a
device \verb+dev+, as well as a ``classical function'' that calls \verb+circuit+ and computes a
cost. The functions can be depicted as nodes in a directed acyclic \emph{computational graph} that
represents the flow of information in the computation. Each node may involve a number of input and
output variables represented by the incoming and outgoing edges, respectively. A
\verb+GradientDescentOptimizer+ is created that improves the initial candidate for these variables
by one step, with the goal of decreasing the cost. PennyLane is able to automatically determine the
gradients of all nodes --- even if the computation is performed on quantum hardware --- and can
therefore compute the gradient of the final cost node with respect to any input variable.

PennyLane is an open-source software project. Anyone who contributes significantly to the library
(new features, new plugins, etc.) will be acknowledged as a co-author of this whitepaper. The source
code for PennyLane is available online on
GitHub\footnote{\url{https://github.com/PennyLaneAI/pennylane/}}, while comprehensive documentation and
tutorials are available on PennyLane.ai\footnote{\url{https://pennylane.ai}}.

In the following, we will introduce the concept of hybrid optimization and discuss how gradients of
quantum nodes are computed. We then present PennyLane's user interface through examples of
optimization and supervised learning, and describe how to write new plugins that connect PennyLane
to other quantum hardware and simulators.

\section{Hybrid optimization}

The goal of optimization in PennyLane is to find the minima of a cost function that quantifies the
quality of a solution for a certain task. In hybrid quantum-classical optimization, the output of
the cost function is a result of both classical and quantum processing, or a \emph{hybrid
computation}. We call the processing submodules \emph{classical} and \emph{quantum nodes}. Both
classical and quantum nodes can depend on tunable parameters $\vars$ that we call \emph{variables},
which are adjusted during optimization to minimize the cost. The nodes can receive inputs $\inp$
from other nodes or directly from the global input to the hybrid computation, and they produce
outputs $f(\inp ; \vars)$. The computation can therefore be depicted as a Directed Acyclic Graph
(DAG) that graphically represents the steps involved in computing the cost, which is produced by the
final node in the DAG. By traversing the DAG, information about gradients can be accumulated via the
rules of automatic differentiation \cite{maclaurin2016modeling, baydin2017automatic}. This is used
to compute the gradient of the cost function with respect to all variables in order to minimize the
cost with a gradient-descent-type algorithm.

\subsection{Quantum nodes}

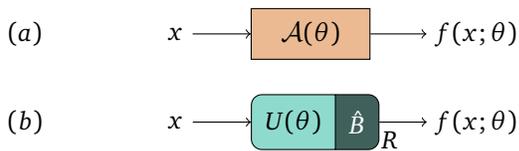
\begin{figure}[t]
    \centering

    \begin{tikzpicture}
        \node (a) at (-2, 0) {$(a)$}; \node[input node] (i) at (0, 0) {$\inp $}; \node[classical
        node] (c) at (1.0,0) {\phantom{A}$\mathcal{A}(\vars )  $\phantom{A}}; \node[input node] (o)
        at (4.0, 0) {$f(\inp ; \vars )$}; \draw[->] (i) -- (c) {}; \draw[->] (c) -- (o);
        \end{tikzpicture}\\ \bigbreak

    \begin{tikzpicture}
        \node (b) at (-2, 0) {$(b)$}; \node[input node] (i) at (0, 0) {$\inp $}; \node[quantum node]
        (c) at (1.0,0) {$U(\vars ) $ \nodepart{two} $\hat{B} $}; \node[input node] (o) at (4.0, 0)
        {$f(\inp ; \vars )$}; \draw[->] (c) -- (o) node[samples label] {$R$}; \draw[->] (i) -- (c);
        \end{tikzpicture}\\
    \bigbreak

    \caption{While a classical node consists of a numerical computation $\mathcal{A}$, a quantum
    node executes a variational circuit~$U$ on a quantum device and returns an estimate of the
    expectation value of an observable~$\hat{B}$, estimated by averaging $R$~measurements.}
    \label{Fig:nodes}
\end{figure}

While classical nodes (see Fig.~\ref{Fig:nodes}(a)) can contain any numerical computations\footnote{
Of course, in order to differentiate the classical nodes the computations have to be based on
differentiable functions.}, quantum nodes have a more restricted layout. A quantum node (in
PennyLane represented by the \verb+QNode+ class) is an encapsulation of a function $f(\inp ; \vars
): \mathbb{R}^m \to \mathbb{R}^n$ that is executed by means of quantum information processing on a
\emph{quantum device}. The device can either refer to quantum hardware or a classical simulator.

\subsubsection{Variational circuits}

The quantum device executes a parametrized quantum circuit called a \emph{variational circuit}
\cite{mcclean2016theory} that consists of three basic operations:
\begin{enumerate}
    \item Prepare an initial state (here assumed to be the  ground or vacuum state $\ket{0}$).
    \item Apply a sequence of unitary gates~$U$ (or more generally, quantum operations or  channels)
    to $\ket{0}$. Each gate is either a fixed operation, or it can depend on some of the
    inputs~$\inp $ or the variables~$\vars $. This prepares the final state $U(\inp, \vars)
    \ket{0}$.
    \item Measure $m$~mutually commuting scalar observables~$\hat{B}_i$ in the final state.
\end{enumerate}

Step 2 describes the way inputs $\inp $ are encoded into the variational circuit, namely by
associating them with gate parameters that are not used as trainable variables\footnote{This
\emph{input embedding} can also be interpreted as a feature map that maps $\inp $ to the Hilbert
space of the quantum system \cite{schuld2018quantum}.}. Step 3 describes the way quantum information
is transformed back to the classical output of a quantum node as the expectated values of the
measured observables:
\begin{equation}
  f_i(\inp; \vars) = \expval{\hat{B}_i} = \bra{0} U(\inp; \vars )^{\dagger}  \hat{B}_i U(\inp; \vars ) \ket{0}.
\end{equation}
The observables $\hat{B}_i$ typically consist of a local observable for each wire (i.e., qubit or
qumode) in the circuit, or just a subset of the wires. For example, $\hat{B}_i$ could be the Pauli-Z
operator for one or more qubits.

\subsubsection{Estimating the expectation values}
\newcommand{\BB}{f_i}
The expectation values $\langle \hat{B}_i\rangle$ are estimated by averaging the measurement results
obtained over $R$ runs of the circuit. This estimator, denoted $\estimator{\BB}$, is unbiased,
$\expval{\estimator{\BB}} = f_i(\inp; \vars)$, and it has variance
\begin{equation}
    \Var{\estimator{\BB}} = \frac{\Var{\hat{B}_i}}{R} = \frac{\expval{\hat{B}_i^2} -\expval{\hat{B}_i}^2}{R},
\end{equation}
which depends on the variance of the operator $\hat{B}_i$, as well as the number of measurements
(`shots') $R$. Note that setting $R=1$ estimates the expectation value from a single measurement
sample. Simulator devices can also choose to compute the exact expectation value numerically (in
PennyLane this is the default behavior, represented by setting \verb+shots=None+). The refined graphical
representation of quantum nodes is shown in Fig.~\ref{Fig:nodes}(b). We will drop the index $R$ in
the following.

\subsubsection{Circuit architectures}

\begin{figure}[t]
\centering
\includegraphics[width=0.45\textwidth]{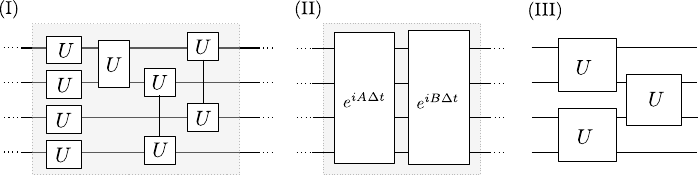}
\caption{Different types of architectures for variational circuits: (I) layered gate architecture,
(II) alternating operator architecture \cite{farhi2014quantum}, and (III) an example of a tensor
network architecture \cite{huggins2018towards}.}
\label{Fig:architectures}
\end{figure}

The heart of a variational circuit is the \emph{architecture}, or the fixed gate sequence that is
the skeleton of the algorithm. Three common types of architectures are sketched in Fig.
\ref{Fig:architectures}. The strength of an architecture varies depending on the desired use-case,
and it is not always clear what makes a good ansatz. Investigations of the expressive power of
different approaches are also ongoing \cite{du2018expressive}. One goal of PennyLane is to
facilitate such studies across various architectures and hardware platforms.

\subsection{Examples of hybrid optimization tasks}

\begin{figure}[t]
    \begin{flushleft}
    \begin{tikzpicture}
        \node[align=left, anchor=west] at (0,1) {(a) Variational quantum eigensolver}; \node[quantum
        node] (q1) at (0,0) {$U(\theta_1) $ \nodepart{two} $\sigma_x$}; \node[quantum node] (qK) at
        (0,-1.5) {$U(\theta_2)$ \nodepart{two} $\sigma_y$}; \node[classical node] (c) at
        (3.25,-0.75) {$(a_1 f_1 + a_2 f_2)^2$}; \node[output node] (o) at (7,-0.75)
        {$\expval{H}^2$}; \draw[] (q1.east) -- (2.75,0) node[out label] {$f_1$} ; \draw[->] (2.75,0)
        -- (2.75,-0.75)-- (c); \draw[] (qK.east) -- (2.75,-1.5) node[out label] {$f_2$}; \draw[->]
        (2.75,-1.5) -- (2.75,-0.75)-- (c); \draw[->] (c) -- (o); \end{tikzpicture}\\ \bigbreak

    \begin{tikzpicture}
        \node[align=left, anchor=west] at (-0.3,1) {(b) Variational quantum classifier}; \node[input
        node] (i) at (0,0) {$\inp $}; \node[classical node] (c1) at (0.75,0) {$\mathcal{P} $};
        \node[quantum node] (q) at (2,0) {$ U(\theta_{W}) $ \nodepart{two} $\sigma_z$} ;
        \node[classical node] (c2) at (5,0) {$ f + \theta_{b}$}; \node[output node] (o) at (7,0)
        {y}; \draw[->] (i) -- (c1); \draw[->] (c1) -- (q); \draw[->] (q) -- (c2) node[out label]
        {$f$}; \draw[->] (c2) -- (o); \end{tikzpicture}\\ \bigbreak

    \begin{tikzpicture}
        \node[align=left, anchor=west] at (-0.3,0.5) {(c) Quantum generative adversarial network
        (QGAN)}; \node[quantum node] (t) at (0,-0.75) {$ D(\theta_D) \circ R$ \nodepart{two}
        $\sigma_z$}; \node[quantum node] (g) at (0,-2.75) {$ D(\theta_D) \circ G(\theta_G)$
        \nodepart{two} $\sigma_z$}; \node[classical node] (p1) at (3.5,-0.75) {$\mathcal{P}_R$};
        \node[classical node] (p2) at (3.5,-2.75) {$\mathcal{P}_G$}; \node[classical node] (c1) at
        (5.,-0.75) {$f_G - f_R$}; \node[classical node] (c2) at (5.,-2.75) {$-f_G$}; \node[output
        node] (o1) at (7.25,-0.75) {$\text{Cost}_D$}; \node[output node] (o2) at (7.25,-2.75)
        {$\text{Cost}_G$}; \draw[->] (t) -- (p1); \draw[->] (g) -- (p2); \draw[->] (p1) -- (c1)
        node[out label] {$f_R$}; \draw[->] (p2) -- (c2) node[out label] {$f_G$}; \draw[->] (p2) --
        (c1) node[out label, left] {$f_G$}; \draw[->] (c1) -- (o1); \draw[->] (c2) -- (o2);
        \end{tikzpicture}\\ \bigbreak

    \end{flushleft}
    \caption{DAGs of hybrid optimization examples. These models and more are available as worked
    examples in the PennyLane docs \cite{pennylane}.}
    \label{Fig:examples}
\end{figure}
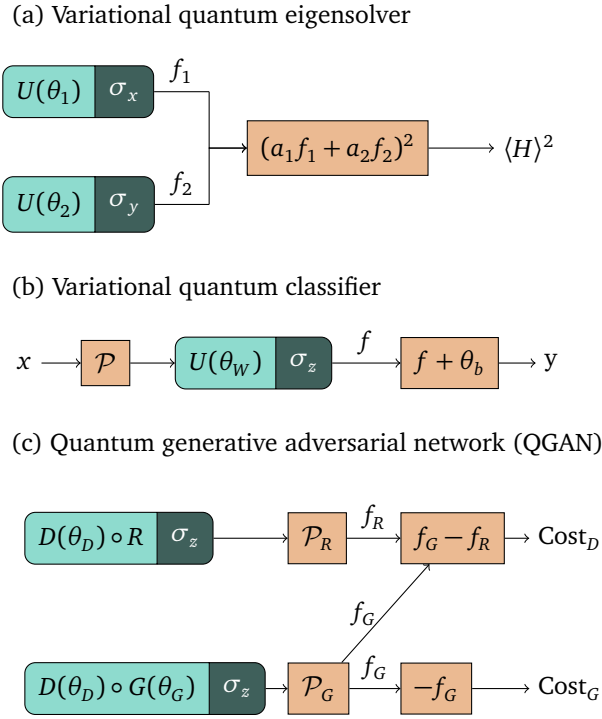

Fig.~\ref{Fig:examples} shows three examples of hybrid optimization tasks depicted as a DAG. Each of
these models is available as a worked example in the PennyLane documentation \cite{pennylane}.
Fig.~\ref{Fig:examples}(a) illustrates a variational quantum eigensolver, in which expectation
values of two Pauli operators are combined with weights $a_1, a_2$ to return the squared global
energy expectation $\expval{H}^2$. Fig.~\ref{Fig:examples}(b) shows a variational quantum classifier
predicting a label $y$ given a data input $x$ for a supervised learning task. The input is
preprocessed by a routine $\mathcal{P}$ and fed into a variational circuit with variables
$\theta_W$. A classical node adds a bias variable $\theta_b$ to the Pauli-Z expectation of a
designated qubit. In Fig.~\ref{Fig:examples}(c) one can see a quantum generative adverserial network
(QGAN) example. It consists of two variational circuits. One represents the ``real data'' circuit
$R$ together with a discriminator circuit $D$, and the other has a ``fake'' generator circuit $G$
replacing $R$. The result is postprocessed by $\mathcal{P}_{R,G}$ and used to construct the cost
function of the discriminator as well as the generator. The goal of a GAN is to train the
discriminator and generator in an adversarial fashion until the generator produces data that is
indistinguishable from the true distribution.

\section{Computing gradients}

PennyLane focuses on optimization via gradient-based algorithms, such as gradient descent and its
variations. To minimize the cost via gradient descent, in every step the individual variables $\mu
\in \Theta$ are updated according to the following rule:\\
\begin{algorithmic}[1]
\Procedure{Gradient Descent step}{} \For {$\mu \in \Theta $} \State $\mu^{(t+1)} = \mu^{(t)} -
\eta^{(t)} \partial_{\mu} C(\Theta)$ \EndFor \EndProcedure
\end{algorithmic}
The learning rate $\eta^{(t)}$ can be adapted in each step, depending either on the step number, or
on the gradient itself. \\

\subsection{Backpropagating through the graph}

A step of gradient descent requires us to compute the gradient $\nabla_{\Theta} C(\Theta)$ of the
cost with respect to all variables $\Theta$. The gradient consists of partial derivatives
$\partial_{\mu} C(\Theta)$ with respect to the individual variables $\mu \in \Theta$. In modern
machine learning libraries like \emph{TensorFlow} \cite{abadi2016tensorflow}, \emph{PyTorch}
\cite{paszke2017automatic}, \emph{Autograd} \cite{maclaurin2015autograd}, or \emph{JAX}
\cite{jax2018github}, this computation is performed using \emph{automatic differentiation}
techniques such as the backpropagation algorithm. PennyLane extends these capabilities to
computations involving quantum nodes, allowing computational models in these four machine learning
libraries (including those with GPU- and TPU-accelerated components) to seamlessly include quantum nodes.
This makes PennyLane completely compatible with standard automatic differentiation techniques
commonly used in machine learning.

While the backpropagation method --- a classical algorithm --- can resolve the gradient
of quantum nodes executed on backpropagation-compatible simulators (such as PennyLane's built-in
\texttt{default.qubit} simulator), this approach suffers from several drawbacks. Firstly,
it does not scale with simulations requiring increasing number of qubits, due to the significant
memory requirements of backpropagation (namely, that the quantum state must be cached
at every step in the simulation). Secondly, it does not support quantum hardware.

To rectify this issue, while preserving the ability to backpropagate through the overall hybrid
quantum-classical computation, note that the backpropagation algorithm does not need to resolve the
quantum information inside quantum nodes --- it is sufficient for us to (separately) compute the
vector Jacobian product of quantum nodes with respect to their (classical) inputs and variables. The
key insight is to use the \emph{same quantum device} (hardware or simulator) that implements a
quantum node to also compute (simulator efficient or hardware compatible) gradients or Jacobians of
that quantum node.

Assume that only the node $n^*$ depends on the subset of variables $\vars\subseteq\Theta$, and that
$\mu$ is in $\vars$. Let $C \circ n^{(p)}_1 \circ \dots \circ n^*$ be the path through the DAG of
(quantum or classical) nodes that emerges from following the cost in the opposite direction of the
directed edges until we reach node $n^*$. Since there may be $N_p \geq 1$ of those paths (see
Fig.~\ref{Fig:dag}), we use a superscript $p$ to denote the path index. All branches that do not
lead back to $\theta$ are independent of $\mu$ and can be thought of as constants. The chain rule
prescribes that the derivative with respect to the variable $\mu \in \theta$ is given
by\footnote{While $\partial_{\mu} C(\Theta)$ is a partial derivative and one entry of the gradient
vector $\nabla C(\Theta)$, intermediate DAG nodes may map multiple inputs to multiple outputs. In
this case, we deal with $2$-dimensional Jacobian matrices rather than gradients.}
\[ \partial_{\mu} C(\Theta) = \sum_{p=1}^{N_p} \frac{\partial C}{\partial n_1^{(p)}} \frac{ \partial
n^{(p)}_1}{\partial n^{(p)}_{2}}  \cdots \frac{\partial n^*}{\partial \mu}. \] In conclusion, we
need to be able to compute two types of gradients for each node: the derivative $\frac{ \partial
n^{(p)}_{i}}{\partial n^{(p)}_{i-1}}$ with respect to the input from a previous node, as well as the
derivative with respect to a node variable $\frac{\partial n}{\partial \mu}$.

\begin{figure}[t]
    \begin{tikzpicture}
        \node[] at (-0.75,1.5) {Path p=1}; \node[classical node, fill=lightgray, anchor=center] (c1)
        at (0,0) {\phantom{.}}; \node[quantum node, anchor=center] (q1) at (1,0) {\phantom{.}
        \nodepart{two} \phantom{.}}; \node[quantum node, fill=lightgray, anchor=center] (q2) at
        (1,0.75) {\phantom{.} \nodepart{two} \phantom{.}}; \node[classical node, anchor=center] (c2)
        at (2,0) {\phantom{.} }; \node[anchor=center] (mid1up) at (2.5,0.325) {};
        \node[anchor=center] (mid1down) at (2.5,-0.325) {}; \node[quantum node, anchor=center] (q3)
        at (3.5,0.75) {\phantom{.} \nodepart{two} \phantom{.}}; \node[classical node,
        fill=lightgray, anchor=center] (c3) at (3,-0.75) {\phantom{.} }; \node[classical node,
        fill=lightgray, anchor=center] (c4) at (4,-0.75) {\phantom{.} }; \node[anchor=center]
        (mid2up) at (4.5,0.325) {}; \node[anchor=center] (mid2down) at (4.5,-0.325) {};
        \node[classical node, anchor=center] (c5) at (5,0) {\phantom{.} }; \node[anchor=center]
        (mid3up) at (5.5,0.325) {}; \node[anchor=center] (mid3down) at (5.5,-0.325) {}; \node[output
        node, anchor=center] (o2) at (6,0) {$C(\Theta)$ }; \draw[->, gray] (c1) -- (q1); \draw[->]
        (q1) -- (c2); \draw[->, gray] (q2) -- (q3); \draw[->, gray] (c2) -| (mid1down)|- (c3);
        \draw[->] (c2) -| (mid1up) |- (q3); \draw[->] (c3) -- (c4); \draw[->, gray] (c4) -| (mid2up)
        |- (c5); \draw[->] (q3) -| (mid2down) |- (c5); \draw[->] (c5) -- (o2); \node[classical node,
        anchor=center] (c5) at (5,0) {\phantom{.} }; \end{tikzpicture}\\

    \begin{tikzpicture}
        \node[] at (-0.75,1.5) {Path p=2}; \node[classical node, fill=lightgray, anchor=center] (c1)
        at (0,0) {\phantom{.}}; \node[quantum node, anchor=center] (q1) at (1,0) {\phantom{.}
        \nodepart{two} \phantom{.}}; \node[quantum node, fill=lightgray, anchor=center] (q2) at
        (1,0.75) {\phantom{.} \nodepart{two} \phantom{.}}; \node[classical node, anchor=center] (c2)
        at (2,0) {\phantom{.} }; \node[anchor=center] (mid1up) at (2.5,0.325) {};
        \node[anchor=center] (mid1down) at (2.5,-0.325) {}; \node[quantum node, fill=lightgray,
        anchor=center] (q3) at (3.5,0.75) {\phantom{.} \nodepart{two} \phantom{.}}; \node[classical
        node, anchor=center] (c3) at (3,-0.75) {\phantom{.} }; \node[classical node, anchor=center]
        (c4) at (4,-0.75) {\phantom{.} }; \node[anchor=center] (mid2up) at (4.5,0.325) {};
        \node[anchor=center] (mid2down) at (4.5,-0.325) {}; \node[classical node, anchor=center]
        (c5) at (5,0) {\phantom{.} }; \node[anchor=center] (mid3up) at (5.5,0.325) {};
        \node[anchor=center] (mid3down) at (5.5,-0.325) {}; \node[output node, anchor=center] (o2)
        at (6,0) {$C(\Theta)$ }; \draw[->, gray] (c1) -- (q1); \draw[->] (q1) -- (c2); \draw[->,
        gray] (q2) -- (q3); \draw[->] (c2) -| (mid1down)|- (c3); \draw[->, gray] (c2) -| (mid1up) |-
        (q3); \draw[->] (c3) -- (c4); \draw[->] (c4) -| (mid2up) |- (c5); \draw[->] (q3) -|
        (mid2down) |- (c5); \draw[->] (c5) -- (o2); \node[classical node, anchor=center] (c5) at
        (5,0) {\phantom{.} };

    \end{tikzpicture}
    \caption{Example illustration of the two paths that lead from the cost function back to a quantum node.}
    \label{Fig:dag}
\end{figure}

\subsection{Derivatives of quantum nodes}

PennyLane provides multiple methods for computing derivatives of quantum nodes with respect to a
variable or input\footnote{When we speak of derivatives here, we actually refer to estimates of
derivatives that result from estimates of expectation values. Numerically computed derivatives in
turn are approximations of the true derivatives, even if the quantum nodes were giving exact
expectations (e.g., by using a classical simulator device).}: hardware-compatible circuit
transforms, backpropagation (if supported by the underlying simulator), or device-provided. By
default, PennyLane uses various heuristics to determine the `best' gradient method --- that is, the
most accurate and efficient gradient method of those supported by the circuit and the device):

\begin{enumerate}
    \item If the device provides its own gradient method, this is the default choice.
    For example, this allows for simulators that support the classical efficient
    'adjoint' method of differentiation \cite{jones2020efficient}.

    \item If the device is computing expectation values exactly (\texttt{shots=None})
    and supports backpropagation, this is the next choice.

    \item Most quantum nodes permit analytic derivatives on hardware via parameter-shift rules
    \cite{schuld2018gradients,wierichs2021general}. If executing on hardware devices, or with
    simulators where \texttt{shots!=None}, a parameter-shift gradient transform is the next best
    choice.

    \item Finally, if the circuit does not permit analytic hardware gradients, numerical
    methods such as the method of finite differences is applied.
\end{enumerate}

\subsubsection{Analytic derivatives}

Recent advances in the quantum machine learning literature \cite{guerreschi2017practical,
farhi2018classification, mitarai2018quantum, schuld2018circuit} have suggested ways to estimate
analytic derivatives by computing linear combinations of different quantum circuits. These rules are
summarized and extended for arbitrary single-frequency operators in a companion paper
\cite{schuld2018gradients}, and extended to operators of arbitrary frequencies in
\cite{wierichs2021general}. This result provides the theoretical foundation for derivative
computations in PennyLane. In a nutshell, PennyLane makes multiple circuit evaluations, taking place
at shifted parameters, in order to compute analytic derivatives. This recipe works for
single-parameter qubit gates of the form $e^{iGx}$, where the Hermitian generator~$G$ has an
equidistant frequency spectrum (which includes e.g., all common qubit parametrized gates), as well
as continuous-variable circuits with Gaussian operations\footnote{For cases that do not fall into
the above two categories, various extensions are available. These include shift rules for gates with
non-equidistant generator frequencies \cite{wierichs2021general}, stochastic parameter-shift rules
for multi-parameter gates \cite{banchi2021measuring}, and a Hadamard test-based approach that
requires an auxiliary qubit \cite{schuld2018gradients}. These extensions are not currently
implemented in PennyLane.}.

If $f(\inp; \vars) = f(\mu)$ is the output of the quantum node, we have
\begin{equation}
  \label{eq:grad_recipe}
  \partial_{\mu} f(\mu) = \sum_{i=1}^r c_i f(\mu + s_i),
\end{equation}
where $r$ is the number of \textit{unique differences} in the eigenvalue spectrum of gate $i$,
$s_i$ the corresponding parameter-shift values, and $c_i$ the coefficients. Note that
$c_i$ and $s_i$ are typically not fixed; there is a degree of freedom that allows the
shift values to be chosen as needed, and the corresponding coefficients to be computed.
Having said that, PennyLane by default chooses shift values that are equidistant with
respect to the gates period, in order to minimize variance.

While this equation bears some structural resemblance to numerical formulas (discussed next), there
are two key differences. First, the values $c_i$ and $s_i$ are not infinitesimal, but finite;
second, Eq.~\eqref{eq:grad_recipe} gives the \emph{exact} derivatives. Thus, while analytic
derivative evaluations are constrained by device noise and statistical imprecision in the averaging
of measurements, they are not subject to numerical issues. To analytically compute derivatives of
qubit gates or gates in a Gaussian circuit, PennyLane automatically computes or looks up the
appropriate derivative recipe for an operation, evaluates the original circuit
multiple times (shifting the argument of the relevant gate by $\{s_i\}$), and takes
the linear combination with coefficients $\{c_i\}$.

\subsubsection{Numerical derivatives}

Numerical derivative methods require only `black-box' evaluations of the model. We estimate the
partial derivative of a node by evaluating its output, $f(\inp; \vars) = f(\mu)$, at several values
which are close to the current value $\mu \in \theta$ ($\mu$ can be either a variable or an input
here). The approximation of the derivative is given by
\begin{equation}
  \partial_{\mu} f(\mu) \approx \frac{f(\mu + \Delta \mu) - f(\mu )}{\Delta \mu}
\end{equation}
for the \emph{forward finite-differences} method, and by
\begin{equation}
\partial_{\mu} f(\mu) \approx \frac{f(\mu + \frac{1}{2} \Delta \mu) - f(\mu - \frac{1}{2}\Delta \mu )}{\Delta \mu}
\label{Eq:findiff}
\end{equation}
for the \emph{centered finite-differences} method. Of course, there is a tradeoff in choice of the
difference $\Delta \mu$ for noisy hardware.

\subsubsection{Backpropagation and device derivatives}
In addition to the analytic and numeric derivative implementations described above --- which are
supported by all simulator and hardware devices --- PennyLane also supports native backpropagation,
as well as directly querying the device for the derivative, if known. For example, a simulator
written using a classical automatic differentiation library, such as TensorFlow, PyTorch, or JAX, can make
use of backpropagation algorithms to calculate derivatives. Compared to the analytic
method on simulators, this may lead to significant time savings, as the information required to
compute the derivative is stored and reused from the forward circuit evaluation --- simply adding
constant overhead. Furthermore, the device derivative may also be used when interfacing with
hardware devices that provide their own custom gradient formulations.

\subsection{Higher-order derivatives}

In addition to computing first-order derivatives of quantum nodes on hardware and simulators,
PennyLane also natively supports arbitrary-order derivatives of quantum nodes. In the case
of gradient transforms that produce multiple circuits to evaluate under-the-hood (such
as the parameter-shift rules and method of finite-differences), the linear combinations
are simply iterated by successively applying the chain and product rules until the required
order is reached. To minimize redundant device evaluations, terms in the iterated rules
are simplified and combined by taking into account the periods of the gates.

\section{User API}

A thorough introduction and review of PennyLane's API can be found in the online documentation. The
documentation also provides several examples for optimization and machine learning of quantum and
hybrid models in both continuous-variable and qubit architectures, as well as tutorials that walk
through the features step-by-step.

\subsection{Optimization}

\begin{figure}[t]
    \begin{center}
        \includegraphics[width=0.7\linewidth]{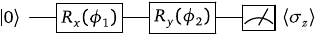}
    \end{center}
    \caption{Variational circuit of the qubit rotation example.}
    \label{Fig:qubit_rot}
\end{figure}

To see how PennyLane allows the easy construction and optimization of variational circuits, let us
consider the simple task of optimizing the rotation of a single qubit --- the PennyLane version of
`Hello world!'.\\

The task at hand is to optimize the variational circuit of Fig.~\ref{Fig:qubit_rot} with two
rotation gates in order to flip a single qubit from state $\ket{0}$ to state $\ket{1}$. After the
rotations, the qubit is in state $\ket{\psi}  = R_y(\phi_2) R_x(\phi_1) | 0 \rangle $ and we measure
the expectation value
\[
f(\phi_1, \phi_2) = \langle \psi | \sigma_z | \psi \rangle = \cos(\phi_1)\cos(\phi_2)
\]
of the Pauli-Z operator. Depending on the variables $\phi_1$ and $\phi_2$, the output expectation
lies between $1$ (if $\ket{\psi} = \ket{0}$) and $-1$ (if $\ket{\psi} = \ket{1}$).\\

PennyLane code for this example --- using the default \emph{autograd} interface for classical
processing --- is shown below in Codeblock~\ref{Cd:rot}. It is a self-contained example that defines
a quantum node, binds it to a computational device, and optimizes the output of the quantum node to
reach a desired target.

\begin{code}
    \pythonfile{code/qubit_rotation.py}
    \caption{Optimizing two rotation angles to flip a qubit.}
    \label{Cd:rot}
\end{code}

We now discuss each element in the above example. After the initial import statements, we declare
the device \verb+dev+ on which we run the quantum node, before defining the quantum node
itself. PennyLane uses the name \verb+wires+ to refer to quantum subsystems (qubits or qumodes)
since they are represented by horizontal wires in a circuit diagram. The decorator
\verb+@qml.qnode(dev)+ is a shortcut that transforms the function \verb+circuit1+ into a quantum
node of the same name. If PennyLane is used with another supported machine learning library, such as
PyTorch or TensorFlow, the \texttt{QNode} interface should be specified when using the decorator,
via the \texttt{interface} keyword argument (\texttt{interface=`torch'} and \texttt{interface=`tf'}
respectively). This allows the QNode to accept objects native to that interface, such as Torch or
TensorFlow tensors.

Note that we could alternatively create the \verb+QNode+ by hand, without the use of the decorator:
\begin{code}
    \begin{minted}{python}
def circuit1():
    ...

circuit1 = qml.QNode(circuit1, dev)
    \end{minted}
    \vspace{0.1cm}
    \caption{Creating a quantum node without the decorator.}
    \label{Cd:create_qnode}
\end{code}
\noindent Finally, the free variables of this computation are automatically optimized through
repeated calls to the \verb+step+ or \verb+step_and_cost+ method of the provided optimizer.

In order for a quantum node to work properly within PennyLane, the function declaring the quantum
circuit must adhere to a few rules. It must contain quantum operations to be applied on the device
(such operations may depend on classical inputs passed to the quantum function), and must return
measurement statistics (including expectation values, variances, and probabilities), of one or more
observables on separate wires. In the latter case, the measurement statistics should be returned
together as a tuple.

\begin{code}
    \pythonfile{code/multi_exp.py}
    \caption{A quantum node that returns two expectations.}
    \label{Cd:rot_sub}
\end{code}
\noindent As long as at least one measurement value is returned, not every wire needs to be
measured. In addition to expectation values, PennyLane also supports returning variances
(\texttt{qml.var()}), probabilities (\texttt{qml.probs()}), and samples (\texttt{qml.sample()}),
although the latter is not differentiable. Simulator devices may also support returning states
(\texttt{qml.state()}, \texttt{qml.density\_matrix()}), which support differentiation on
backpropagation-capable devices. Tensor products of observables may also be specified using the \texttt{@}
notation, for example \texttt{qml.expval(qml.PauliZ(0) @ qml.PauliY(2))}. Finally, Hamiltonians
representing linear combinations of operators can be specified via the notation:
\begin{code}
    \begin{minted}{python}
qml.expval(qml.PauliZ(0) @ qml.PauliY(2) + 0.5 * qml.PauliZ(1))
    \end{minted}
    \vspace{0.1cm}
    \caption{Specifying an expectation value of a Hamiltonian.}
    \label{Cd:hamiltonian}
\end{code}

Multiple quantum nodes can be bound to the same device, and the same circuit can be run on different
devices. In the latter case, the \verb+QNode+ will need to be created manually. These use-cases are
shown in Codeblock \ref{Cd:multiple}.\footnote{This particular example leverages the Qiskit
\cite{mckay2018qiskit} plugin for PennyLane \cite{pennylaneqiskit}. This code will not run without
the plugin being installed and without hardware access credentials being provided.}

\begin{code}
    \pythonfile[firstline=4, firstnumber=1]{code/multiple_devices.py}
    \caption{Constructing multiple quantum nodes from various circuits and devices.}
    \label{Cd:multiple}
\end{code}

\noindent If we have multiple quantum nodes, we can combine the outputs with a classical node to
compute a final cost function:
\begin{code}
    \pythonfile{code/class_node.py}
    \caption{A classical node combining two quantum nodes.}
\end{code}

\noindent This cost compares a simulator and a hardware, and finds values of the variables for which
the two produce the same result. This simple example hints that automatic optimization tools could
be used to correct for systematic errors on quantum hardware.\\

In summary, quantum and classical nodes can be combined in many different ways to build a larger
hybrid computation, which can then be optimized automatically in PennyLane.

\subsection{Supervised learning}

PennyLane has been designed with quantum and hybrid quantum-classical machine learning applications
in mind. To demonstrate how this works, we consider a basic implementation of a variational
classifier. A variational classifier is a model where part of the computation of a class prediction
is executed by a variational circuit. The circuit takes an input $x$ as well as some trainable
variables and computes a prediction $y$.

\begin{code}
    \pythonfile{code/qml.py}
    \caption{Code stub for creating a variational quantum classifier.}
    \label{Cd:qml}
\end{code}

\noindent In Codeblock \ref{Cd:qml}, the machine learning model is defined in the \verb+model+
function. It retrieves two types of variables, a scalar bias and a list of layer
weights. It then computes the output of the variational circuit and adds the bias. The variational
circuit, in turn, first refers to a routine that encodes the input into a quantum state, and then
applies layers of a certain gate sequence, after which an expectation is returned.\\

We can train the classifier to generalize the input-output relation of a training dataset.

\begin{code}
    \pythonfile{code/qml2.py}
    \caption{Code stub for optimizing the variational classifier.}
    \label{Cd:qml2}
\end{code}
\noindent The variables are initialized as a tuple containing the bias and the weight matrix. In the
optimization loop, we feed a Python lambda function into the optimizer. Since the optimizer expects
a function with a single input argument, this is a way to feed both $X$ and $Y$ into the cost.

PennyLane can straightforwardly incorporate various standard machine learning practices. Examples
include: optimizing minibatches of data with stochastic gradient descent, adding more terms to the
cost, saving variables to a file, and continuing optimization with a warm start. For full worked-out
examples, see the PennyLane documentation \cite{pennylane}.\\

\subsection{Behind the scenes}

The core feature of PennyLane that enables such seamless optimization integration is the ability to
easily extract gradients of hybrid quantum-classical cost functions, regardless of underlying
quantum devices. The approach for computing hybrid gradients depends on the autodifferentiation
library of choice; below, we demonstrate this capability using the default Autograd integration, but
the same can be done using PyTorch, TensorFlow, or JAX --- simply use the canonical method of
computing gradients in the chosen autodifferentiation library. In the default Autograd interface,
\verb+qml.grad+ and \verb+qml.jacobian+ compute gradients of classical or quantum nodes. Let us
switch to ``interactive mode'' and look at \verb+circuit1+ and \verb+circuit2+ from above.
\begin{code}
    \pythonfile{code/grad.py}
    \caption{Computing gradients of hybrid functions.}
\end{code}
As expected, the gradient of a \verb+QNode+ with $2$ inputs and $1$ output is a 1-dimensional array,
while the Jacobian of a \verb+QNode+ with $2$ inputs and $2$ outputs is a $2 \times 2$ array. The
\verb+Optimizer+ class uses gradients and Jacobians computed this way to update variables. PennyLane
currently has several built-in optimizers which work with the default Autograd interface. This
includes standard optimization techniques from classical machine learning (standard gradient
descent, gradient descent with momentum, gradient descent with Nesterov momentum, Adagrad, Adam,
RMSprop), as well as a suite of 'quantum aware' optimizers, which take into account the quantum
geometry and hardware to increase convergence while minimizing required quantum resources (quantum
natural gradient descent \cite{stokes2019quantum}, coordinate minimization
\cite{ostaszewski2021structure,wierichs2021general}, shot adpative optimization
\cite{arrasmith2020operator}, and Riemannian gradient-flow \cite{wiersema2022optimizing}). If using
PyTorch, TensorFlow, or JAX, the optimizers provided by those libraries can be used.

\section{Algorithms and features}

PennyLane also provides a higher-level interface for easily and automatically creating and
processing QNodes. This includes a library of circuit ans\"atze or `templates' from across the
quantum machine learning literature, libraries of transforms to manipulate circuits and QNodes, and
the ability to easily create cost functions for common quantum variational algorithms.

\subsection{Templates}

The \texttt{pennylane.templates} module provides a growing library of pre-coded templates of common
variational circuit architectures that can be used to build, evaluate, and train more complex
models. In the literature, such architectures are commonly known as an ansatz. PennyLane
conceptually distinguishes two types of templates, layer architectures and input embeddings. Most
templates are complemented by functions that provide an array of random initial parameters.

\begin{code}
    \begin{minted}{python}
import pennylane as qml
from pennylane import numpy as np

dev = qml.device('default.qubit', wires=2)

@qml.qnode(dev)
def circuit(weights, x):
    qml.AngleEmbedding(x, [0,1])
    qml.StronglyEntanglingLayers(weights, [0,1])
    return qml.expval(qml.PauliZ(0))

shape = qml.StronglyEntanglingLayers.shape(n_layers=3, n_wires=2)
weights = np.random.random(shape, requires_grad=True)
print(circuit(weights, x=[1., 2.]))
    \end{minted}
    \vspace{0.1cm}
    \caption{The embedding template \texttt{AngleEmbedding} is used to embed data within the QNode,
    and the layer template {StronglyEntanglingLayers} used as the variational ansatz with a uniform
    parameter initialization strategy.}
\end{code}
Templates provided include \texttt{AmplitudeEmbedding}, \texttt{QAOAEmbedding},
\texttt{CVNeuralNetLayers}, among others. In addition, custom templates can be easily created; simply
create a quantum function that applies quantum gates:
\begin{code}
    \begin{minted}{python}
def bell_state_preparation(wires):
    qml.Hadamard(wires=wires[0])
    qml.CNOT(wires=wires)
    \end{minted}
    \vspace{0.1cm}
    \caption{Defining a custom template.}
\end{code}
\noindent The custom template can then be used within any valid QNode.

\subsection{Transforms}

While the ability to define, process, execute, and train quantum nodes enables the design of rich
variational models, the power of differentiable quantum programming with PennyLane is fully unlocked
by `transforms'; a library of functions that manipulate, transform, and extract
information from quantum functions.

There are two main forms of transforms available in PennyLane:

\begin{enumerate}
    \item \textbf{Classical transforms}: These transforms extract information from quantum functions
    without execting the underlying device. Examples include \texttt{qml.draw()} (for drawing
    quantum circuits), \texttt{qml.specs()} (resource information), and \texttt{qml.matrix()}
    (extract the matrix representation of the circuit unitary).

    \item \textbf{Quantum transforms}: These transforms extract information from quantum nodes by
    generating one or more quantum circuits, and post-processing the results with classical
    processing.
\end{enumerate}
Thus, in contrast to the `classical' transform, the quantum transform requires additional
quantum device evaluations in order to compute the requested quantity. Aside from this
conceptual difference, both forms of transforms share the same three important qualities:
\begin{itemize}
    \item They take as input a \textit{function}, and transform it to a \textit{new} function
    that takes the same arguments, but returns a different quantity.
    \item If the output of the transformed function is one or more floating point values that
    depends smoothly on the input to the original function (as with \texttt{qml.matrix}), then the
    transformed function is typically differentiable with respect to the function arguments.
    \item If the transform itself permits floating point parameters, then the
    transformed function is typically differentiable with respect to the transform arguments.
\end{itemize}
More formally, we can define a differentiable quantum transform as follows:
\begin{defn}
    Let $f(\theta_i)$ be a quantum function with input parameters $\{\theta_i\}$.
    A transform $\mathcal{T}$ with inputs $\{\phi_i\}$ is a differentiable quantum transform if
    \begin{align}
        \mathcal{T}(f)\rightarrow \{g_k\},
    \end{align}
    where each $g_k$ is also a differentiable quantum function with respect to the same inputs
    $\{\theta_i\}$, and $\frac{\partial \mathcal{T}}{\partial \phi_{i}}$ is defined for all
    $\{\phi_i\}$.
  \end{defn}
Such transforms are common-place in PennyLane, with examples being the parameter-shift rules,
and \texttt{qml.batch\_inputs()} which transforms a circuit to permit batched input embedding.
Another example is the transform that returns the Fubini-Study metric tensor of a quantum
node:
\begin{code}
    \pythonfile{code/metric_tensor.py}
    \caption{Differentiating a transformed QNode.}
\end{code}
Circuit compilation is another example of a quantum transform, and is in fact a special case,
as compilation transforms always map a quantum function to a \textit{single} output
quantum function. As such, they can be arbitrarily composed and `stacked'. PennyLane provides
a variety of compilation transforms:
\begin{code}
    \pythonfile{code/compile.py}
    \caption{Applying compilation transforms.}
\end{code}
In addition, the \texttt{qml.compile()} transform makes it easy to build custom compilation
pipelines from these individual transform building blocks.

Finally, PennyLane provides tools for creating custom transforms. These work by manipulating
the low-level datastructure representing a sequence of operations and measurements
to be executed on the quantum device --- the quantum \textit{tape}. Two decorators
are available: \texttt{qml.qfunc\_transform}, for defining transforms that map a quantum
function to a single quantum function, and \texttt{qml.batch\_transform}, for defining
transforms that map a quantum function to multiple quantum functions.

\begin{code}
    \begin{minted}{python}
@qml.batch_transform
def my_transform(tape, *transform_params):
    ...
    return new_tapes, processing_fn
    \end{minted}
    \vspace{0.1cm}
    \caption{Batch transforms take an input tape representing a quantum circuit, and return
    a list of tapes to execute on the device, as well as a classical post-processing function
    to apply to the execution results.}
\end{code}

\subsection{Just-in-time compilation}

In addition to providing quantum transforms, PennyLane also continues to work
with many of the composable functional transforms available via autodifferentiation
frameworks. One example includes just-in-time (JIT) compilation, the ability to dynamically
compile components of the computation to machine code, enabling both performance improvements,
and the ability to execute on resources such as GPUs and TPUs. JAX and TensorFlow both provide
JIT transformations --- \texttt{jax.jit} and \texttt{tf.function} respectively --- that
transform wrapped functions to enable JIT compatibility. These transformations work seamlessly
with cost functions that include quantum nodes, regardless of where the quantum node is executed.
Noteably, this allows models to be constructed that take advantage of JIT compilation to speed
up classical pre- and post-processing, while retaining the ability to execute quantum components
on quantum hardware.

\begin{code}
    \begin{minted}{python}
import pennylane as qml
import jax
from jax import numpy as jnp

s3_bucket = ("my-bucket", "my-prefix")

dev = qml.device(
    "braket.aws.qubit", 40,
    "arn:aws:braket:::device/qpu/rigetti/Aspen-11",
    s3_bucket
)

@qml.qnode(dev, interface="jax")
def circuit(x):
    qml.RX(x[0], wires=0)
    qml.RY(x[1], wires=1)
    qml.CNOT(wires=[0, 1])
    return qml.expval(qml.PauliZ(0))

@jax.jit
def cost(x):
    return jnp.abs(1 - circuit(jnp.sin(x)))

x = jnp.array([-0.5, 0.6])
cost(x)
    \end{minted}
    \vspace{0.1cm}
    \caption{A hybrid classical-quantum cost function is just-in-time compiled using JAX. The
    contained QNode is executed on quantum hardware using the Amazon Braket plugin.}
\end{code}
In addition, there is support for differentiation transforms, including \texttt{jax.grad},
\texttt{jax.jacobian}, \texttt{jax.vjp}, \texttt{torch.autograd.functional}, and TensorFlow's
\texttt{tape.jacobian} and \texttt{tape.gradient}.

\subsection{Quantum chemistry}
The variational quantum eigensolver (VQE) algorithm is frequently applied to quantum chemistry
problems \cite{peruzzo2014variational}. In VQE, a quantum computer is first used to prepare the
trial wave function of a molecule, and the expectation value of its electronic Hamiltonian is
measured. A classical optimizer then adjusts the quantum circuit parameters to find the lowest
eigenvalue of the Hamiltonian.

The starting point of VQE is an electronic Hamiltonian expressed in the Pauli basis --- however,
determining the Pauli-basis representation from the molecular structure is highly non-trivial,
requiring use of both self-consistent field methods as well as mapping of Fermionic states and
operators to qubits. PennyLane provides a quantum chemistry package that, with a single line of
code, can be used to generate the electronic Hamiltonian of a molecule. It employs an in-built
fully-differentiable Hartree-Fock solver~\cite{arrazola2021differentiable}, in addition to
supporting external quantum chemistry packages such as OpenFermion~\cite{mcclean2017openfermion},
PySCF~\cite{sun2018pyscf}, and Psi4~\cite{turney2012psi4,parrish2017psi4}.

To build the Hamiltonian, it is necessary to specify the atomic symbols and the geometry of the molecule. Additional optional information includes the charge of the molecule, the spin-multiplicity of the
Hartree-Fock state, and the atomic basis set. The following example
code generates the qubit Hamiltonian for the neutral hydrogen molecule using the \texttt{sto-3g}
basis set for atomic orbitals:

\begin{code}
    \pythonfile{code/qchem.py}
    \caption{Generating the electronic Hamiltonian of the Hydrogen molecule.}
\end{code}

Once the Hamiltonian has been generated, a circuit is constructed and standard PennyLane techniques
are used to optimize the circuit parameters:

\begin{code}
    \begin{minted}{python}
dev = qml.device("default.qubit", wires=qubits)
opt = qml.GradientDescentOptimizer(stepsize=0.4)

hf = qml.qchem.hf_state(electrons=2, orbitals=4)

@qml.qnode(dev)
def circuit(parameters):
    qml.BasisState(hf, wires=range(qubits))
    qml.DoubleExcitation(parameters[0], wires=[0, 1, 2, 3])
    return qml.expval(H)

params = np.zeros(1, requires_grad=True)

prev_energy = 0.0
for n in range(50):
    params, energy = opt.step_and_cost(circuit, params)
    print(energy, params)
    if np.abs(energy - prev_energy) < 1e-6:
        break
    prev_energy = energy
    \end{minted}
    \vspace{0.1cm}
    \caption{Constructing a VQE optimization workflow.}
    \label{Cd:plugin:vqe}
\end{code}

\section{Built-in simulator devices}

While PennyLane is designed to easily integrate with external quantum devices (see
\hyperref[sec:plugin]{Writing a plugin device} for more details), it also includes a suite
of built-in simulators, to allow for immediate exploration of differentiable quantum
programming without needing to install additional dependencies. This
allows for a rapid-iteration style of development; explore the capabilities of the
quantum algorithm under development, before scaling it up to run on quantum hardware.

Currently, PennyLane provides five simulator devices:

\begin{itemize}
    \item \texttt{default.qubit}: A Python-based qubit statevector simulator, with backends written
    using NumPy, TensorFlow, PyTorch, and JAX. As a result, this simulator supports
    end-to-end backpropagation, and models containing this device can be deployed for execution
    on GPUs and TPUs. Due to the memory overhead of backpropagation, this device works best
    for 0-20 qubits.
    \item \texttt{default.mixed}: A Python-based qubit mixed-state simulator, written in NumPy.
    Allows for quantum nodes that contain quantum channels.
    \item \texttt{default.gaussian}: A Python-based continuous-variable simulator, written using
    NumPy, and designed to support photonic-based quantum nodes. This device supports
    continuous-variable quantum circuits with Gaussian gates and measurements.
    \item \texttt{lightning.qubit}: A high-performance qubit statevector simulator,
    written in C++. This device supports the adjoint method of quantum differentiation
    \cite{jones2020efficient}, enabling extremely efficient optimization for quantum nodes
    with 20 or more qubits.
    \item \texttt{lightning.gpu}: A high-performance qubit statevector simulator,
    written using NVIDIA's cuQuantum SDK~\cite{cuquantum} for GPU accelerated circuit simulation.
    As with \texttt{lightning.qubit}, adjoint differentiation is supported.
\end{itemize}

\subsection{\texttt{lightning.qubit} and \texttt{lightning.gpu}}

As \texttt{default.qubit} provides an easy way to explore the use of PennyLane, often more involved
workflows require a high-performance backend; \texttt{lightning.qubit} was developed with this in
mind. The core functionality is written using modern C++ language features (11, 14, and 17), and allows
for an extensible implementation of quantum gate kernels.

While most users may be running on x86-64 systems, \texttt{lightning.qubit} also provides pre-built
support for ARM and PowerPC platforms allowing us to target all architectures for our users,
from laptops to cloud and HPC systems. As we provide pre-built wheels for \texttt{lightning.qubit}, this
will be automatically be installed alongside PennyLane, without any need for user compilation.

\texttt{lightning.qubit} is both designed for optimal performance on the individual kernel level, as
well as high throughput jobs that are common to PennyLane: namely, differentiable workflows of
quantum circuits. Our implementation of the adjoint differentiation method directly parallelizes
over user-requested observables, and offers signficant run-time improvements for workloads with many
observable evaluations. As a result, \texttt{lightning.qubit} using adjoint differentiation
significantly reduces the time-to-solution over other simulators and gradient methods. 

As an extension to \texttt{lightning.qubit}, we also provide \texttt{lightning.gpu}, where gate calls are offloaded
to the NVIDIA cuQuantum SDK. By taking advantage of the additional performance provided by GPUs,
\texttt{lightning.gpu} can evaluate gradients of much larger and deeper quantum circuits that
would otherwise have been intractible on CPU resources alone.

\section{Writing a plugin device}\label{sec:plugin}

PennyLane was designed with extensibility in mind, providing an API for both hardware devices and
software simulators to easily connect and allow PennyLane access to their frameworks. This enables
the automatic differentiation and optimization features of PennyLane to be used on an external
framework with minimal effort. As a result, PennyLane is inherently hardware agnostic --- the user
is able to construct hybrid computational graphs containing \verb+QNode+s executed on an arbitrary
number of different devices, and even reuse quantum circuits across different devices. As of version
0.24, PennyLane has plugins available for the Xanadu Cloud and Strawberry Fields
\cite{killoran2018strawberry, pennylanesf,arrazola2021quantum}, Amazon Braket \cite{braket}, Rigetti
\cite{smith2016practical,pennylaneforest}, IBM Quantum and Qiskit
\cite{mckay2018qiskit,pennylaneqiskit}, Google Cirq \cite{pennylanecirq}, ProjectQ
\cite{steiger2018projectq, pennylanepq}, Microsoft QDK \cite{pennylaneqdk}, Qulacs
\cite{qulacs,pennylanequlacs}, AQT \cite{pennylaneaqt}, Honeywell \cite{pennylanehoneywell}, and
IonQ \cite{pennylaneionq}.

In PennyLane, there is a subtle distinction between the terms `plugin' and `device':
\begin{itemize}
    \item A plugin is an external Python package that provides additional quantum \textit{devices}
    to PennyLane.
    \item Each plugin may provide one (or more) devices, that are accessible directly by PennyLane,
    as well as any additional private functions or classes.
\end{itemize}
Once installed, these devices can be loaded directly from PennyLane without any additional steps
required by the user. Depending on the scope of the plugin, a plugin can also provide custom quantum
operations, observables, and functions that extend PennyLane --- for example by converting from the
target framework's quantum circuit representation directly to a QNode supporting
autodifferentiation\footnote{One example being the PennyLane-Qiskit plugin, which provides
conversion functions \texttt{qml.from\_qasm()} and \texttt{qml.from\_qiskit()} --- allowing QNodes
to be created from QASM and Qiskit quantum programs respectively.}. In the remainder of this
section, we briefly describe the plugin API of PennyLane, and how it can be used to provide new
quantum devices.

\subsection{Devices}

When performing a hybrid computation using PennyLane, one of the first steps is to specify the
quantum devices which will be used by quantum nodes. As seen above, this is done as follows:
\begin{code}
    \begin{minted}{python}
import pennylane as qml
dev1 = qml.device(short_name, wires=2)
    \end{minted}
    \vspace{0.1cm}
    \caption{Loading a PennyLane-compatible device.}
    \label{Cd:plugin:device}
\end{code}
\noindent where \texttt{short\_name} is a string which uniquely identifies the device provided. In
general, the short name has the following form: \texttt{pluginname.devicename}.

\subsection{Creating a new device}

The first step in making a PennyLane plugin is creating the device class. This is as simple as
importing the abstract base class \texttt{Device} from PennyLane, and subclassing it \footnote{See
the developers guide in the PennyLane documentation,
\href{https://pennylane.readthedocs.io/en/stable/development/plugins.html}
{https://pennylane.readthedocs.io/en/stable/development/plugins.html},
for an up-to-date guide on creating a new plugin}:

\begin{code}
\begin{minted}{python}
from pennylane import Device

class MyDevice(Device):
    """MyDevice docstring"""
    name = 'My custom device'
    short_name = 'example.mydevice'
    pennylane_requires = '0.1.0'
    version = '0.0.1'
    author = 'Ada Lovelace'
\end{minted}
    \vspace{0.1cm}
    \caption{Creating a custom PennyLane-compatible device.}
    \label{Cd:plugin:makedevice}
\end{code}

Here, we have begun defining some important class attributes (`identifiers') that allow PennyLane to
recognize the device. These include:
\begin{itemize}
    \item \texttt{Device.name}: a string containing the official name of the device
    \item \texttt{Device.short\_name}: the string used to identify and load the device by users of
    PennyLane
    \item \texttt{Device.pennylane\_requires}: the version number(s) of PennyLane that this device
    is compatible with; if the user attempts to load the device on a different version of PennyLane,
    a \texttt{DeviceError} will be raised
    \item \texttt{Device.version}: the version number of the device
    \item \texttt{Device.author}: the author of the device
\end{itemize}
Defining all these attributes is mandatory.

\subsection{Supporting operations and expectations}

Plugins must inform PennyLane about the operations and expectations that the device supports, as
well as potentially further capabilities, by providing the following class attributes/properties:

\begin{itemize}
    \item \texttt{Device.operations}: a set of the supported PennyLane operations as strings, e.g.,
    \texttt{operations = \{"CNOT", "PauliX"\}}. This is used to decide whether an operation is
    supported by your device in the default implementation of the public method
    \texttt{Device.supported()}.
    \item \texttt{Device.observables}: a set of the supported PennyLane observables as strings,
    e.g., \texttt{observables = \{"PauliX", "Hadamard", "Hermitian"\}}. This is used to decide
    whether an observable is supported by your device in the default implementation of the public
    method \texttt{Device.supported()}.
    \item \texttt{Device.\_capabilities}: a dictionary containing information about the capabilities
    of the device. For example, the key \verb+'model'+, which has value either \verb+'qubit'+ or
    \verb+'CV'+, indicates to PennyLane the computing model supported by the device. This class
    dictionary may also be used to return additional information to the user --- this is accessible
    from the PennyLane frontend via the public method \texttt{Device.capabilities}.
\end{itemize}

A subclass of the \texttt{Device} class, \texttt{QubitDevice}, is provided for easy integration with
simulators and hardware devices that utilize the qubit model. \texttt{QubitDevice} provides
automatic support for all supported observables, including tensor observables. For a better idea of
how these required device properties work, refer to the two reference devices.

\subsection{Applying operations and measuring statistics}

Once all the class attributes are defined, it is necessary to define some required class methods, to
allow PennyLane to apply operations to your device. In the following examples, we focus on the
\texttt{QubitDevice} subclass. When PennyLane evaluates a \verb+QNode+, it calls the
\texttt{Device.execute} method, which performs the following process:

\begin{code}
    \begin{minted}{python}
self.check_validity(circuit.operations, circuit.observables)

# apply all circuit operations
self.apply(circuit.operations, rotations=circuit.diagonalizing_gates)

# generate computational basis samples
if (not self.analytic) or circuit.is_sampled:
    self._samples = self.generate_samples()

# compute the required statistics
results = self.statistics(circuit.observables)

return self._asarray(results)
    \end{minted}
    \vspace{0.1cm}
    \caption{The PennyLane \texttt{Device.execute} method, called whenever a quantum node is
    evaluated.}
\end{code}

In most cases, there are a minimum of two methods that need to be defined:
\begin{itemize}
    \item \texttt{Device.apply}: Accepts a list of PennyLane Operations to be applied. The
    corresponding quantum operations are applied to the device, the circuit rotated into the
    measurement basis, and, if relevant, the quantum circuit compiled and executed.
    \item \texttt{Device.probability}: Returns the (marginal) probability of each computational
    basis state from the last run of the device.
\end{itemize}
In addition, if the device generates/returns its own computational basis samples for measured modes
after execution, the following method must also be defined:
\begin{itemize}
    \item \texttt{Device.generate\_samples}: Generate computational basis samples for all wires. If
    \texttt{Device.generate\_samples} is not defined, PennyLane will automatically generate samples
    using the output of the device probability.
\end{itemize}
Once the required methods are defined, the inherited methods \texttt{Device.expval},
\texttt{Device.var}, and \texttt{Device.sample} can be passed an observable (or tensor product of
observables), returning the corresponding measurement statistic.

\subsection{Installation and testing}

PennyLane uses a \texttt{setuptools} \texttt{entry\_points} approach to plugin integration. In order
to make a plugin accessible to PennyLane, the following keyword argument to the \texttt{setup}
function must be provided in the plugin's \texttt{setup.py} file:

\begin{code}
    \begin{minted}{python}
devices_list = [ 'myplugin.mydev1 = MyMod.MySubMod:MyDev1', 'myplugin.mydev2 = MyMod.MySubMod:MyDev2'
]
setup(entry_points={'pennylane.plugins': devices_list})
    \end{minted}
    \vspace{0.1cm}
    \caption{Creating the PennyLane device entry points.}
    \label{Cd:plugin:entrypoints}
\end{code}
\noindent Here, \texttt{devices\_list} is a list of devices to be registered,
\texttt{myplugin.mydev1} is the short name of the device, and \texttt{MyMod.MySubMod} is the path to
the Device class, \texttt{MyDev1}. To ensure the device is working as expected, it can be installed
in developer mode using \texttt{pip install -e pluginpath}, where \verb+pluginpath+ is the location
of the plugin. It will then be accessible via PennyLane.

All plugins should come with unit tests, to ensure that devices work as expected. In general, as all
supported operations have their gradient formula defined and tested by PennyLane, testing that the
device calculates the correct gradients is not required --- it is sufficient to test that it
\textit{applies} and \textit{measures} quantum operations and observables correctly. To help,
PennyLane provides a device integration test utility, to ensure that a specified device returns
expected values for various circuits and measurements. This device testing utility comes pre-installed
with PennyLane, and is available via the command \texttt{pl-device-test}. For example,
running the device test against the built-in \texttt{default.qubit} simulator:

\begin{code}
    \begin{minted}{python}
pl-device-test --device default.qubit --shots 10000 --skip-ops
    \end{minted}
    \vspace{0.1cm}
    \caption{Running the PennyLane device integration test suite against \texttt{default.qubit} with
    10000 shots, and skipping the tests of any unsupported operations.}
\end{code}

\subsection{Supporting new operations}

PennyLane also provides the ability to add custom operations or observables to be executed on the
plugin device, that may not be currently supported by PennyLane. For qubit architectures this is
done by subclassing the \texttt{Operation} and \texttt{Observable} classes, defining the number of
parameters the operation takes, and the number of wires the operation acts on. In addition, if the
frequencies of the operation are known, the corresponding \verb+parameter_frequencies+ should be
provided, to open up analytic differentiation support in PennyLane.

For example, to define the U2 gate, which depends on parameters $\phi$ and $\lambda$, we create the
following class:
\begin{code}
    \begin{minted}{python}
class U2(Operation):
    """U2 gate."""
    num_params = 2
    num_wires = 1
    parameter_frequencies = [(1,), (1,)]

    def __init__(self, phi, lam, wires, **kwargs):
        super().__init__(phi, delta, wires=wires, **kwargs)

    @staticmethod
    def compute_decomposition(phi, lam, wires):
        decomp_ops = [
            Rot(lam, np.pi / 2, -lam, wires=wires),
            PhaseShift(phi + lam, wires=wires)
        ]
        return decomp_ops
    \end{minted}
    \vspace{0.1cm}
    \caption{Creating a custom qubit operation.}
    \label{Cd:plugin:operation}
\end{code}
\noindent where the following quantities \textit{must} be declared:
\begin{itemize}
    \item \texttt{Operation.num\_params}: the number of parameters the operation takes

    \item \texttt{Operation.num\_wires}: the number of wires the operation acts on
\end{itemize}

In addition, the following optional \textit{operator representations} can be defined,
which enables additional functionality within PennyLane:

\begin{itemize}
    \item \texttt{Operation.compute\_matrix}: static method which returns the matrix representation
    in the computational basis.

    \item \texttt{Operation.compute\_sparse\_matrix}: static method which returns the sparse matrix representation
    in the computational basis.

    \item \texttt{Operation.compute\_decomposition}: static method which returns a list of operators
    representing the tensor product decomposition.

    \item \texttt{Operation.compute\_diagonalizing\_gates}: static method which returns a list of PennyLane
    operations that diagonalize the observable in the computational basis.

    \item \texttt{Operation.compute\_eigvals}: static method which returns the eigenvalues.

    \item \texttt{Operation.compute\_kraus\_matrices}: static method which returns the a list of
    Kraus matrices representing a channel.

    \item \texttt{generator}: An instance method that returns an operator representing the
    Hermitian generator of a single parameter operation.

    \item \texttt{Operation.parameter\_frequencies}: property or attribute that defines the
    frequency spectrum of an operator with respect to an expectation value. If provided, this
    is used to compute generalized shift rules for the operator, enabling anlaytic quantum
    gradients on hardware.

    \item \texttt{Operation.grad\_recipe}: the gradient recipe for operation.
    This is a list with one tuple per operation parameter. For parameter $k$, the tuple is of the
    form $(c_k, m_k, s_k)$, resulting in a gradient recipe of $$\frac{d}{d\phi_k}O =
    \sum_k c_k O(m_k \phi_k+s_k).$$

    \item \texttt{Operation.label}: determines how the operation appears in a circuit diagram.
\end{itemize}

The user can then import this operation directly from your plugin, and use it when defining a
\verb+QNode+:
\begin{code}
    \begin{minted}{python}
import pennylane as qml
from MyModule.MySubModule import Ising

@qnode(dev1)
def my_qfunc(phi):
    qml.Hadamard(wires=0)
    Ising(phi, wires=[0, 1])
    return qml.expval(qml.PauliZ(1))
    \end{minted}
    \vspace{0.1cm}
    \caption{Using a plugin-provided custom operation.}
    \label{Cd:plugin:usingoperation}
\end{code}
\noindent In this case, as the plugin is providing a custom operation not supported by PennyLane, it
is recommended that the plugin unit tests \textit{do} provide tests to ensure that PennyLane returns
the correct gradient for the custom operations.

\subsubsection{Custom observables}

Custom observables can be added in an identical manner to operations above, but with three small
changes:

\begin{itemize}
    \item The \texttt{Observable} class should instead be subclassed.
    \item The static method \texttt{Observable.compute\_eigvals} should be defined, returning a
    one-dimensional array of eigenvalues of the observable.
    \item The static method \texttt{Observable.compute\_diagonalizing\_gates} should be defined.
    This is used to support devices that can only perform measurements in the computational basis.
\end{itemize}

\subsubsection{Custom CV operations and expectations}

For custom continuous-variable operations or expectations, the \texttt{CVOperation} or
\texttt{CVObservable} classes must be subclassed instead. In addition, for CV operations with known
analytic gradient formulas (such as Gaussian operations), the static class method
\texttt{CV.\_heisenberg\_rep} must be defined:
\begin{code}
    \begin{minted}{python}
class Custom(CVOperation):
    """Custom gate"""
    n_params = 2
    n_wires = 1
    grad_method = 'A'
    grad_recipe = None

    @staticmethod
    def _heisenberg_rep(params):
        return function(params)
    \end{minted}
    \vspace{0.1cm}
    \caption{Creating a custom continuous-variable operation.}
\end{code}
For operations, the \verb+_heisenberg_rep+ method should return the Heisenberg representation of the
operation, i.e., the matrix of the linear transformation carried out by the operation for the given
parameter values\footnote{Specifically, if the operation carries out a unitary transformation $U$,
this method should return the matrix for the adjoint action $U^\dagger (\cdot) U$.}. This is used
internally for calculating the gradient using the analytic method (\verb+grad_method = 'A'+). For
observables, this method should return a real vector (first-order observables) or symmetric matrix
(second-order observables) of coefficients which represent the expansion of the observable in the
basis of monomials of the quadrature operators. For single-mode operations we use the basis
$\mathbf{r} = (\mathbb{I}, \hat{x}, \hat{p})$, and for multi-mode operations the basis $\mathbf{r} =
(\mathbb{I}, \hat{x}_0, \hat{p}_0, \hat{x}_1, \hat{p}_1, \ldots)$, where $\hat{x}_k$ and $\hat{p}_k$
are the quadrature operators of qumode~$k$. Note that, for every gate, even if the analytic gradient
formula is not known or if \texttt{\_heisenberg\_rep} is not provided, PennyLane continues to
support the finite difference method of gradient computation.

\section{Conclusion}
We have introduced PennyLane, a Python package that extends automatic differentiation to quantum and
hybrid classical-quantum information processing. This is accomplished by introducing a new
\emph{quantum node} abstraction which interfaces cleanly with existing DAG-based automatic
differentiation methods like the backpropagation algorithm. The ability to compute gradients of
variational quantum circuits -- and to integrate these seamlessly as part of larger hybrid
computations -- opens up a wealth of potential applications, in particular for optimization and
machine learning tasks.

We envision PennyLane as a powerful tool for many research directions in quantum computing and
quantum machine learning, similar to how libraries like TensorFlow, PyTorch, or JAX have become
indispensible for research in deep learning. With small quantum processors becoming publicly
available, and with the emergence of variational quantum circuits as a new algorithmic paradigm, the
quantum computing community has begun to embrace heuristic algorithms more and more. This spirit is
already common in the classical machine learning community and has -- together with dedicated
software enabling rapid exploration of computational models -- allowed that field to develop at a
remarkable pace. With PennyLane, tools are now freely available to investigate model structures,
training strategies, and optimization landscapes within hybrid and quantum machine learning, to
explore existing and new variational circuit architectures, and to design completely new algorithms
by circuit learning.

\bibliography{references}

\end{document}